\pdfoutput=1 
\documentclass[fleqn,10pt]{wlscirep}
\usepackage{pdfpages}
\usepackage{graphicx,hyperref} 
\usepackage{caption}
\usepackage{color}
\usepackage{stackengine}
\usepackage[position=t,singlelinecheck=off]{subcaption}
\usepackage{array}
\usepackage{multirow}
\usepackage{enumerate}
\usepackage{amsmath}
\usepackage{amssymb}
\usepackage{amsthm}
\usepackage{amsfonts}
\usepackage{amssymb}
\usepackage{mathtools}
\usepackage{url}
\usepackage{bm}
\usepackage[utf8]{inputenc}
\usepackage[T1]{fontenc}
\usepackage[scaled]{helvet}%
\usepackage{lmodern}
\usepackage{microtype} 
\usepackage{lineno} 
\usepackage{xcolor}
\usepackage[normalem]{ulem}
\definecolor{mygreen}{RGB}{76, 175, 80}
\definecolor{myblue}{RGB}{33,150,243}
\definecolor{myred}{RGB}{255, 87, 34}

\makeatletter
\newcommand*{\rom}[1]{\expandafter\@slowromancap\romannumeral #1@}
\makeatother
\newcommand\given[1][]{\:#1\vert\:}
\hypersetup{
  colorlinks   = true, 
  urlcolor     = black, 
  linkcolor    = blue, 
  citecolor   = blue 
}
\def\qmod{q^{\text{mod}}}
\def\conf{{\text{conf}}}
\def\qint{q^{\text{int}}}
\def\qexp{q^{\text{exp}}}
\def\qcnd{q^{\text{cnd}}}
\def\vol{\text{vol}}

\definecolor{purple}{RGB}{170, 0, 255}
\title{A generalised significance test for individual communities in networks}
\author[1,2]{Sadamori Kojaku}
\author[2,*]{Naoki Masuda}
\affil[1]{CREST, JST, Kawaguchi Center Building, 4-1-8, Honcho, Kawaguchi-shi, Saitama 332-0012, Japan}
\affil[2]{Department of Engineering Mathematics,
Merchant Venturers Building, University of Bristol,
Woodland Road, Clifton, Bristol, BS8 1UB, United Kingdom}
\affil[*]{naoki.masuda@bristol.ac.uk}

\begin{abstract}
Many empirical networks have community structure, in which nodes are densely interconnected within each community (i.e., a group of nodes) and sparsely across different communities. 
Like other local and meso-scale structure of networks, communities are generally heterogeneous in various aspects such as the size, density of edges, connectivity to other communities and significance. 
In the present study, we propose a method to statistically test the significance of individual communities in a given network. 
Compared to the previous methods, the present algorithm is unique in that it accepts different community-detection algorithms and the corresponding quality function for single communities.
The present method requires that a quality of each community can be quantified and that community detection is performed as optimisation of such a quality function summed over the communities. 
Various community detection algorithms including modularity maximisation and graph partitioning meet this criterion. 
Our method estimates a distribution of the quality function for randomised networks to calculate a likelihood of each community in the given network. 
We illustrate our algorithm by synthetic and empirical networks.
\end{abstract}
\begin{document}

\flushbottom
\maketitle
%
%
\thispagestyle{empty}


\section{Introduction}
\label{sec:introduction}
Many biological, physical and social systems can be expressed as networks, with nodes representing individual entities within the network and edges representing pairwise relationships between nodes \cite{Newman2010,Barabasi2016}.
Among various structural properties of networks, many empirical networks have community structure such that a network is composed of communities, which are groups of nodes that are densely interconnected with each other while sparsely interconnected with those in other groups \cite{Fortunato2010,Fortunato2016}. 
A community may correspond to the role of nodes. 
For example, communities may correspond to functional modules of proteins \cite{Jonsson2006}, groups of airports serving the same geographical region \cite{Guimera2005} and herds of people sharing an interest \cite{Newman2006}. 

Many algorithms have been proposed for finding communities in networks \cite{Fortunato2010,Fortunato2016}.
These algorithms are often equipped with a quality function with which to judge whether or not the detected community structure is significant overall. 
A much less asked fundamental question is the significance of individual communities.
In fact, a network may be composed of a part where community structure is pronounced and another part where community structure is vague or absent. 
To discuss community structure in such a ``chimera'' network, one needs methods to assess statistical significance of single communities. 

In the present study, we consider the significance of single communities that have been detected by a non-overlapping community-detection algorithm. 
An algorithm for testing significance of individual communities was previously proposed \cite{Lancichinetti2010}.
In that algorithm, one uses a quality function for individual communities to compare the quality of a community in question, detected in the given network, and that detected in randomised networks.
The distribution of the quality function in randomised networks is analytically known.
The authors then used the same significant test in OSLOM, which is an algorithm for finding various types of communities \cite{Lancichinetti2011}. 
However, OSLOM does not optimise the same quality function as that used in the aforementioned statistical test or its aggregate over the different communities. 
The same discrepancy exists in a different significance test for single communities \cite{Spirin2003}. 
In an extreme case, let us suppose one detects communities by optimising a quality function that is very different from the quality function used in the statistical test. Then, the detected communities may have small values of the quality function used in the statistical test and will be judged to be insignificant.
However, in terms of the quality function used in the community detection, the detected communities may be sufficiently strong.

This pitfall may be overcome if one uses the same quality function for the community detection and the statistical test.
There exist such significance tests for individual communities \cite{Wang2008,Zhao2011}.
However, these significance tests \cite{Wang2008,Zhao2011} do not consider the possible dependence of the quality function value on the size of community \cite{Spirin2003,Leskovec2010,Yang2015}. 
This practice is problematic for the following reason. 
Suppose that two communities in the given network have different sizes and bear the same value of the quality function. 
Then, the significance level (i.e., $p$-value) in these statistical tests is the same for the two communities. 
In general, however, the quality function value may be positively correlated with the community size, which is in fact often the case (Section \ref{sec:quality_and_size}).  
In this case, it is easier for the larger community to attain the observed quality function value than for the smaller community under the null model. 
Then, the smaller community should be judged to be more significant than the larger community if they yield the same quality function value.
An aforementioned statistical test does consider the dependence of the quality function value on the community size \cite{Spirin2003}. 
However, that method does not use a common quality function between community detection and statistical testing, as discussed already.

Based on these considerations, it will be useful to develop methods to test the significance of individual communities that (i) use a quality function that is consistent with the one used in community detection, and (ii) take into account the dependence of the quality function value on the community size. 
We will develop a new statistical test for individual communities that meets these criteria.
An additional feature of our method is that it allows for general quality functions.
Python code for the present significance test is available at \href{https://github.com/skojaku/qstest/}{\textcolor{blue}{https://github.com/skojaku/qstest/}}.

\section{Methods}
\label{sec:methods}
\subsection{Correlation between quality and community size}
\label{sec:quality_and_size}
We consider unweighted networks composed of $N$ nodes. Denote their $N \times N$ adjacency matrix by ${\bm A}=(A_{ij})$, where $A_{ij} = 1$ if nodes $i$ and $j$ are adjacent and $A_{ij}=0$ otherwise.
We assume that the network is undirected (i.e., $A_{ij}=A_{ji}$ for all $i \neq j$) and does not contain self-loops (i.e., $A_{ii}=0$).
Let $M$ be the number of edges in the network. 
We denote by $d_i \equiv \sum_{j=1} ^N A_{ij}$ the degree of node $i$. 

\begin{figure}
    \includegraphics[width=0.95\hsize]{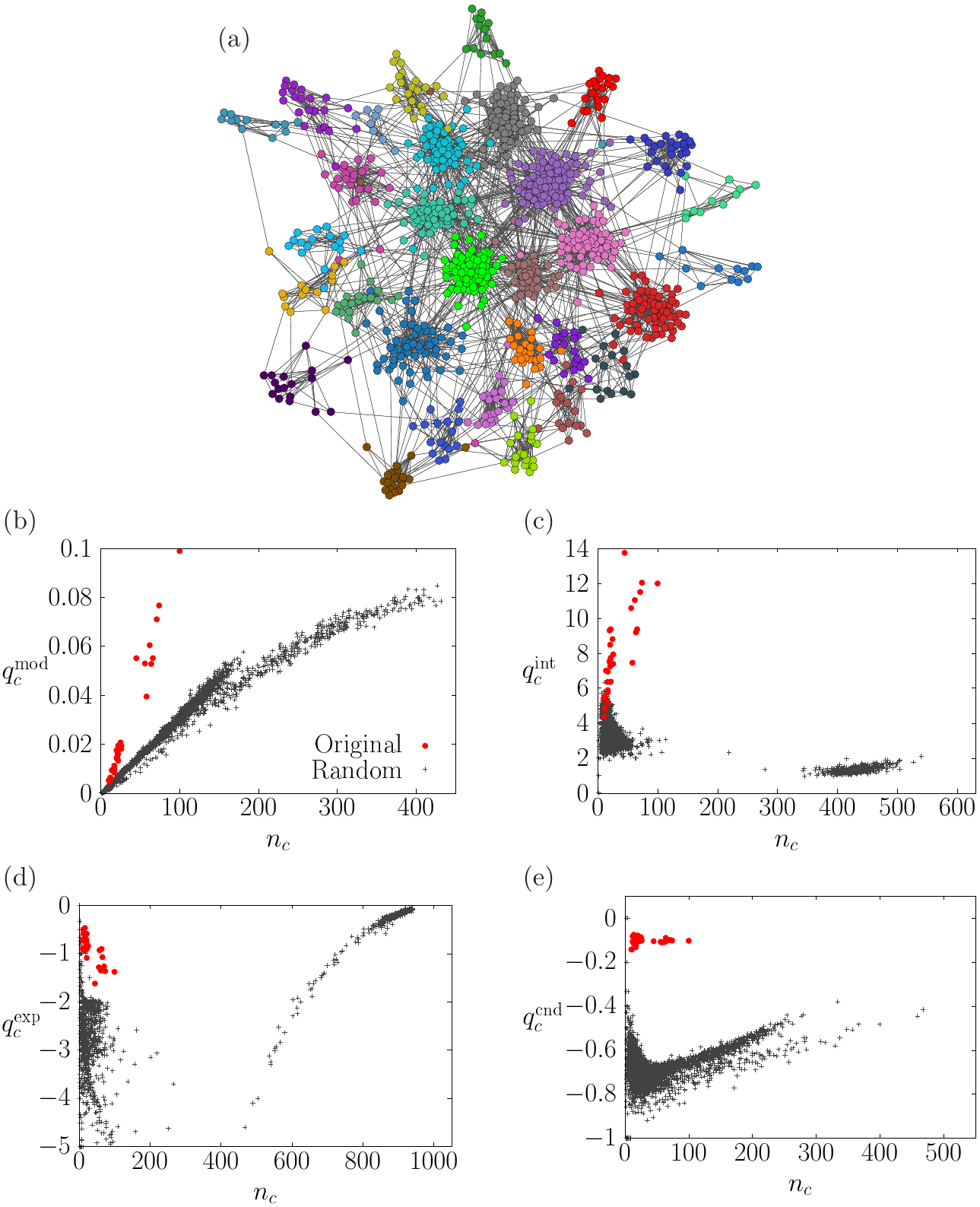}
    \caption{
        (a) A network with 31 non-overlapping communities generated by the LFR model.
        The circles represent nodes. 
        The lines between the nodes represent edges. 
        The colour of each node indicates the planted community to which the node belongs.   
        (b)--(e) Quality of a community (i.e., $\qmod_c$, $\qint_c$, $\qexp_c$ and $\qcnd_c$) plotted against its number of nodes, $n_c$.
        The circles indicate the planted communities shown in panel (a).
        The crosses indicate the communities detected in 500 randomised networks generated by the configuration model. 
        To find communities in the randomised networks, we use the Louvain algorithm \cite{Blondel2008} for $\qmod_c$ (panel (b)) and 
        a variant of the Kernighan--Lin algorithm (Section~\ref{sec:community_detection}) for $\qint _c$, $\qexp _c$ and $\qcnd _c$  (panels (c)--(e)).
    }
    \label{fig:lfr}
\end{figure}

One may regard a community as significant if its quality value is significantly larger than that expected for randomised networks.
This intuitive approach has a problem. 
To see this, let us consider a benchmark network generated by the Lancichinetti-Fortunato-Radicchi (LFR) model \cite{Lancichinetti2008} (Fig.~\ref{fig:lfr}(a)).  
The network has $N=10^3$ nodes and consists of $C$ non-overlapping communities. 
Each node $i$ belongs to one of the $C=31$ communities.
To generate the network, we set the average node's degree to 10, the maximum node's degree to 100, the range of the number of nodes in a community $c$ (denoted by $n_c$) to $[10, 100]$ and the power-law exponent for the distributions of $d_i$ and $n_c$ to 2. 
Let us consider a quality function $\qmod_c$ given by \cite{Leskovec2010,Yang2015}  

\begin{linenomath}
\begin{align}
    \label{eq:qmod}
     \qmod_c\equiv \frac{1}{2M}\sum_{\substack{1 \leq i,j \leq N\\ i,j \in \text{community $c$}}} \left( A_{ij} - \frac{d_id_j}{2M}\right).
\end{align}
\end{linenomath}
Note that the modularity is the sum of $\qmod_c$ over the communities \cite{Newman2006}.
We find a strong positive correlation between $\qmod_c$ and $n_c$ (circles in Fig.~\ref{fig:lfr}(b)). 
This is also true for communities in randomised networks that are generated by the configuration model, i.e., random networks that preserve the expected degree of each node (crosses in Fig.~\ref{fig:lfr}(b)).
Crucially, large communities detected in the randomised networks have larger $\qmod_c$ values than small communities in the original network do. 
Therefore, we can not judge the significance of communities solely by the value of $\qmod_c$.
The results are qualitatively the same for other quality functions for individual communities introduced in the following section (Figs. \ref{fig:lfr}(c), \ref{fig:lfr}(d), and \ref{fig:lfr}(e)).

\subsection{Our statistical test}
On the basis of the observations made in the previous section, we construct a statistical test for individual communities as follows.
Note that we do not specify the quality function $q_c$, which may be $\qmod_c$ or a different one.
Moreover, we do not specify how one measures the size  $s_c$ of community $c$. 
We refer to the present statistical test based on a quality function $q$ and community size $s$ as the $(q, s)$--test. 

Suppose that we have a community $c$ with quality $q_c$ and size $s_c$. 
We judge community $c$ to be significant if its $q_c$ value is larger than those for communities of the same size $s_c$ detected in randomised networks. 
We compute  $P(\tilde q \geq q_c \given s_c )$, which is the probability that a community of size $s_c$ detected in randomised networks generated by the configuration model has a quality value $\tilde q$ larger than $q_c$.
We numerically estimate $P(\tilde q \geq q_c \given s_c)$ as follows.
First, we generate 500 randomised networks using the configuration model.  
Then, we detect communities in each randomised network by the algorithm that has been used to detect communities in the original network.
Let $\overline C$ be the sum of the number of communities detected in the $500$ randomised networks.
For each community $\overline{c}$ ($1 \leq \overline{c} \leq \overline{C}$) in the randomised networks, we compute the quality $\tilde q_{\overline{c}}$ and size $\tilde s_{\overline{c}}$.
Then, we compute the average values, i.e., $\langle \tilde q \rangle \equiv \sum_{{\overline{c}}=1}^{\overline{C}} \tilde q_{\overline{c}} / \overline{C}$ and $\langle \tilde s \rangle \equiv \sum_{{\overline{c}}=1}^{\overline{C}} \tilde s_{\overline{c}} / \overline{C}$, and 
the unbiased estimation of the standard deviation, i.e., $\sigma_{\tilde q} \equiv \sqrt{\sum_{{\overline{c}}=1} ^{\overline{C}}  (\tilde q_{{\overline{c}}} -\langle \tilde q\rangle )^2 / (\overline{C}-1)}$ and $\sigma_{\tilde s} \equiv \sqrt{\sum_{{\overline{c}}=1} ^{\overline{C}}  (\tilde s_{{\overline{c}}} -\langle \tilde s\rangle )^2 / (\overline{C}-1)}$.
We estimate the joint probability distribution $P(\tilde q, \tilde s)$ using the kernel density estimator \cite{Wand1993} as follows: 
\begin{linenomath}
\begin{align}
    \label{eq:joint}
    P(\tilde q, \tilde s) &= \left.\sum_{{\overline{c}}=1} ^{\overline{C}} f\left( \frac{\tilde q - \tilde q_{{\overline{c}}}}{h\sigma_{\tilde q}}, \frac{\tilde s - \tilde s_{{\overline{c}}} }{h\sigma_{\tilde s}} \right) \middle/  \overline{C} \right.,
\end{align}
\end{linenomath}
where $h$ is the width of the kernel.
The function $f(\cdot, \cdot)$ is the bivariate Gaussian kernel (i.e., bivariate standard normal distribution) given by  
\begin{linenomath}
\begin{align}
    \label{eq:bivariate}
    f(x_1,x_2) \equiv \frac{1}{2\pi \sqrt{1-\gamma ^2}} \exp\left( -\frac{ x_1 ^2  - 2 \gamma x_1 x_2 + x_2 ^2 }{2\left(1-\gamma^2 \right)}  \right), 
\end{align}
\end{linenomath}
where   
\begin{linenomath}
\begin{align}
    \label{eq:pearson}
    \gamma \equiv \frac{
            \sum_{{\overline{c}}=1}^{\overline{C}} \left( \tilde q_{{\overline{c}}} - \langle \tilde q\rangle \right)\left( \tilde s_{{\overline{c}}} - \langle \tilde s\rangle  \right) 
            }{ 
                \sqrt{\sum_{{\overline{c}}=1}^{\overline{C}} \left( \tilde q_{{\overline{c}}} - \langle \tilde q\rangle \right)^2}
                \sqrt{\sum_{{\overline{c}}=1}^{\overline{C}} \left( \tilde s_{{\overline{c}}} - \langle \tilde s\rangle \right)^2}
            },  
\end{align}
\end{linenomath}
is the Pearson correlation coefficient between $\{ \tilde q_{{\overline{c}}} \}_{{\overline{c}}=1} ^{\overline{C}} $ and $\{ \tilde s _{{\overline{c}}} \}_{{\overline{c}}=1} ^{\overline{C}} $.
The probability distribution estimated by the Gaussian kernels is close to any form of the true probability distribution as the number of samples increases \cite{Parzen1962}.
Although there are also non-Gaussian kernels that share this property \cite{Parzen1962}, we used the Gaussian kernels, which is a state-of-the-art method.
The width $h$ is a free parameter that affects the speed of the convergence to the true probability distribution.
Optimising the value of $h$ requires assumptions for the true probability distributions and intensive computations \cite{Park1990,Jones1996}.
Therefore, we set $h=\overline{C}^{(-1/6)}$ according to Scott's rule-of-thumb \cite{Scott2012}, which often provides a reasonable estimate in practice \cite{Park1990,Jones1996,Scott2012}.

The conditional probability, $P(\tilde q > q_c \given s_c)$, is given by 
\begin{linenomath}
\begin{align}
    \label{eq:pval0}
    P(\tilde q \geq q_c \given s_c) &=  \dfrac{\displaystyle \int ^{\infty} _{q_c} P( \tilde q, s_c ) {\rm d}\tilde q}{\displaystyle \int ^{\infty} _{-\infty} P(\tilde q, s_c)  {\rm d}\tilde q}
    =  \dfrac{\displaystyle \sum\limits_{\overline{c}=1}^{\overline{C}} \int ^{\infty} _{q_c} f\left( \frac{\tilde q - \tilde q_{\overline{c}}}{\sigma_{\tilde q}h}, \frac{ s_c - \tilde s _{{\overline{c}}}}{\sigma_{\tilde s}h} \right) {\rm d}\tilde q}{\displaystyle \sum\limits_{{\overline{c}}=1}^{\overline{C}} \int ^{\infty} _{-\infty} f\left( \frac{\tilde q - \tilde q_{{\overline{c}}}}{\sigma_{\tilde q}h}, \frac{s_c - \tilde s _{{\overline{c}}}}{\sigma_{\tilde s}h} \right) {\rm d}\tilde q}.    
\end{align}
\end{linenomath}
The integration of $f(x_1, x_2)$ over $x_1$ yields  
\begin{align}
    \label{eq:bivariatecum}
    \int_{y} ^{\infty} f\left( x_1 , x_2 \right){\rm d}x_1 = \frac{1}{\sqrt{2\pi }}\exp\left(-\frac{x_2 ^2}{2}\right)\left[ 1  -\Phi\left( \frac{ y -\gamma x_2 }{\sqrt{1-\gamma^2}} \right)\right],
\end{align}
where $\Phi\left( \cdot \right)$ is the cumulative distribution function of the standard normal distribution.
By substituting Eq.~\eqref{eq:bivariatecum} into Eq.~\eqref{eq:pval0}, we have 
\begin{linenomath}
\begin{align}
    \label{eq:pval}
    P(\tilde q \geq q_c \given s_c) &= 1 - \dfrac{
                    \displaystyle \sum\limits_{{\overline{c}}=1}^{\overline{C}}{ 
                        \exp\left[ -\left(\frac{s_c -  \tilde s_{{\overline{c}}} }{\sqrt{2}h\sigma_{\tilde s}}\right)^2\right]}
                    \Phi\left(
            \frac{1}{\sqrt{1-\gamma^2}} \left( \frac{q_c  - \tilde q_{{\overline{c}}}}{h\sigma_{\tilde q}} - \gamma \frac{s_c  - \tilde s_{{\overline{c}}}}{h\sigma_{\tilde s}}\right) 
            \right)
                    }{
                    \displaystyle \sum\limits_{{\overline{c}}=1}^{\overline{C}}{\exp\left[ -\left(\frac{s_c -  \tilde s_{{\overline{c}}} }{\sqrt{2}h\sigma_{\tilde s}}\right)^2\right]}
                    }. 
\end{align}
\end{linenomath}
Finally, we regard community $c$ as significant if $P(\tilde q \geq q_c \given s_c) \leq \alpha$, where $\alpha \in [0,1]$ is the significance level.
The conditional probability $P(\tilde q \geq q_c \given s_c)$ obeys a uniform probability distribution over $[0,1]$ for a community detected in a randomised network (see Supplementary Information 1).
One can estimate more accurate $p$-values (i.e. $P(\tilde q \geq q_c \given s_c)$) using a larger number of randomised networks, which, however, requires an additional computational time.
We opt to use $500$ randomised networks to obtain sufficiently accurate $p$-values in a reasonable time.
In fact, the $p$-value does not change much if one increases the number of randomised networks beyond 500 or if one uses networks with different numbers of nodes and communities (Supplementary Information 2).

As the number of communities, $C$, increases, some insignificant communities would be significant owing to the multiple comparison problem.
To avoid this, we use the {\v{S}}id{\'{a}}k correction \cite{Sidak1967}, i.e.,  
$\alpha = 1-(1-\alpha')^{1/C}$, where $\alpha' \in [0,1]$ is the targeted significance level.
We set $\alpha' = 0.05$.

\subsection{Time complexity}
\label{sec:timecomplexity}
The time complexity of the proposed statistical test is evaluated as follows.
Generating one randomised network from the configuration model consumes ${\cal O}(N + M)$ time using an efficient algorithm \cite{Miller2011}, which is implemented in some network analysis software \cite{Networkit,Networkx}.
For each generated randomised network, we detect communities. 
Any community-detection algorithm qualified for the present statistical test computes the quality and size of the individual communities and maximises the quality function for the entire network.
We use the quality and size of the optimised communities in the statistical test.
We carry out these procedures for each of the $R$ randomised networks, consuming ${\cal O}((N+M+Z)R)$ time in total, where $Z$ is the time complexity of the community-detection algorithm.
We compute the $p$-value for each of the $C$ communities in the original network using Eq.~\eqref{eq:pval} with $RC^{\conf}$ samples on average, $\{ \tilde q_{c} \}_{c = 1} ^{RC^\conf}$ and $\{ \tilde s_{c} \}_{c=1} ^{RC^\conf}$, where $C^{\conf}$ is the average number of communities detected in a randomised network.
This incurs a time complexity of ${\cal O}(C\times RC^{\conf})$.  
In total, the proposed statistical test requires ${\cal O}((N + M + Z + CC^{\conf})R)$ time.

The time complexity can be mitigated using parallel computing.
In other words, one runs multiple threads, each of which generates independent samples of $(\tilde q_{c}, \tilde s_{c})$.
Once the sampling is completed in all the threads, one computes the $p$-value using Eq.~\eqref{eq:pval}.
We used 16 threads on a computer with the Intel 2.6GHz Sandy Bridge processors and 4GB of memory.
For the largest network we analysed (i.e., Internet \cite{KONECT}; $N=34,761$ nodes), our statistical test needed 403 seconds using the Louvain community-detection algorithm, which has a time complexity of ${\cal O}(M)$ \cite{Blondel2008}.
With the Kernighan-Lin community-detection algorithm having a time complexity of ${\cal O}(N^2)$ \cite{Karrer2011}, it took 17,763 seconds (i.e. approximately 5 hours).

\subsection{Community detection with different quality functions}
\label{sec:community_detection}
Among various quality functions for individual communities apart from $\qmod_c$ \cite{Leskovec2010,Yang2015,Fortunato2016}, we consider the following three quality functions.
The internal average degree  \cite{Yang2015} (i.e., normalised number of intra-community edges), denoted by $\qint_c$, is defined by 
\begin{linenomath}
\begin{align}
    \qint _c &\equiv \frac{1}{n_c} \sum \limits _{{\substack{1 \leq i,j \leq N\\ i,j \in \text{community $c$}}}} A_{ij}. \label{eq:qint} 
\end{align}
\end{linenomath}
The maximisation of $\qint _c$ yields a community having dense intra-community connectivity. 
The expansion \cite{Yang2015}, denoted by $\qexp_c$, is defined by  
\begin{linenomath}
\begin{align}
    \qexp _c &\equiv - \frac{1}{n_c} \sum \limits _{{\substack{1 \leq i,j \leq N\\ i \in \text{community $c$}\\ j \notin \text{community $c$}}}} A_{ij}. \label{eq:qexp}
\end{align}
\end{linenomath}
The maximisation of $\qexp _c$ yields a community having sparse inter-community connectivity. 
Finally, the conductance \cite{Yang2015}, denoted by $\qcnd_c$, is defined by 
\begin{linenomath}
\begin{align}
    \qcnd _c &\equiv - \frac{1}{\vol_c}  \sum \limits _{{\substack{1 \leq i,j \leq N\\ i \in \text{community $c$}\\ j \notin \text{community $c$}}}} A_{ij}, \label{eq:qcnd}
\end{align}
\end{linenomath}
where $\vol_c$ is the sum of degrees of nodes (i.e., volume) in a community $c$.
Similar to the case of $\qexp_c$, the maximisation of $\qcnd _c$ yields a community having sparse inter-community connectivity. 
One can also interpret the maximisation of $\qcnd_c$ as the maximisation of the number of intra-community edges \cite{Luxburg2007}. 

For $\qmod_c$, we adopt the Louvain algorithm to maximise the modularity (i.e., sum of $\qmod_c$ over the communities, $\sum_{c=1}^C \qmod_c$) to find communities in the original and randomised networks. 
However, the Louvain algorithm is not available to $Q=\sum_{c=1}^C q_c$, where $q_c = \qint_c$, $\qexp_c$ or $\qcnd_c$.
Therefore, we adopt a variant of the Kernighan--Lin algorithm \cite{Kernighan1970} used in a previous study \cite{Karrer2011}.
The algorithm seeks partitioning of the network into communities that maximises $Q$.
Suppose that each node $i$ has a tentative label $\ell_i$ ($1 \leq \ell_i \leq C$) indicating the index of the community to which node $i$ belongs. 
First, we assign each node to one of the $C$ communities selected uniformly at random.
Second, for each node $i$, we tentatively relabel it to a different label and measure the increment in $Q$. 
Third, we select the node $i$ and its new label $c$ that maximise the increment in $Q$ among all nodes $i$ ($1\le i\le N$) and all possible new labels. 
Regardless of whether $Q$ increases or not, we accept the proposed relabelling of node $i$ (i.e., set $\ell_i = c$). 
Fourth, we determine the pair of another node $j$ ($j\neq i$) and its tentative new label $c'$, which maximises the increment in $Q$, change the label of $j$ to $c'$ (i.e., $\ell_j = c'$). 
In this manner, we relabel nodes one by one. Here we do not relabel the nodes that have already been relabelled.
After sequentially relabelling the $N$ nodes, we select the labelling that yields the largest value of $Q$ among the $N+1$ labellings that have appeared in the course of relabelling the $N$ nodes. 
If the initial labelling (before relabelling any node) yields the largest value of $Q$, we terminate the algorithm. 
Otherwise, we use the labelling that has yielded the largest $Q$ value among the $N+1$ labellings as the initial labelling in the next round of updating the labels.
We repeat the aforementioned procedure to sequentially relabel $N$ nodes and select the best labelling.
We repeat rounds of updating until the initial labelling is the best labelling in the round in terms of the $Q$ value.

To find communities in networks using $\qint_c$, $\qexp_c$ or $\qcnd_c$, we need to specify the number of communities, $C$.
Otherwise, the maximisation of the quality functions may yield trivial communities. 
For example, $\qexp_c$ is always the largest when each connected component constitutes a community because there is no inter-community edge. 
In the analysis of synthetic networks, we set $C$ to the number of planted communities. 
For empirical networks, we set $C$ to the number of communities identified by the Louvain algorithm.
 
\subsection{Other statistical tests} 
We compare the $(q,s)$--test with two statistical tests, i.e., the test proposed by Spirin and Mirny \cite{Spirin2003} and the test proposed by Lancichinetti, Radicchi and Ramasco \cite{Lancichinetti2010}, which we refer to as the S--test and L--test, respectively.
As is the case with the $(q,s)$--test, both S--test and L--test adopt the configuration model as the null model. 
For both statistical tests, we set the significance level for a single community to $\alpha = 1 - (1-\alpha')^{1/C}$, where $\alpha' = 0.05$. 

The S--test regards a community as significant if it has more intra-community edges than a community composed of the same number of nodes detected in randomised networks does.
Their original algorithm \cite{Spirin2003} is slow for large networks.
Therefore, we adopt the Kernighan--Lin algorithm \cite{Kernighan1970} to optimise the quality function for a community adopted in the S--test. 
Up to our numerical efforts, our implementation is faster and also finds better community structure than their original algorithm does in terms of their quality function. 
  
The L--test regards a community as significant if every node in the community has more neighbours within the community than that expected for the configuration model.
In the original paper \cite{Lancichinetti2010}, the authors defined two significance measures, i.e., ${\cal C}$-score and ${\cal B}$-score.
We adopt the ${\cal B}$--score, which is less conservative than the ${\cal C}$--score.
In the original article ~\cite{Lancichinetti2010}, the ${\cal B}$--score is claimed to be more trustworthy than the ${\cal C}$--score because 
the ${\cal C}$--score but not the B-score relies on an extreme value statistics.

\subsection{Data}
\label{sec:data}
We apply the statistical test to the 12 empirical networks listed in Table \ref{ta:empnet}.
We ignore the directions and weights of edges in the empirical networks.

The karate club network represents the relationships among the members of a university's karate club \cite{Zachary1977}. 
Each node represents a member of the karate club. 
Two members are defined to be adjacent if they are friends outside of the club activities.

The dolphin social network represents the relationships of the dolphins living near Doubtful Sound in New Zealand \cite{Lusseau2003}. 
Each node represents a dolphin.
Two dolphins are defined to be adjacent if they are frequently observed in the same school. 

The network of Les Mis\'{e}rables represents the relationships between the characters of a novel, Les Mis\`{e}rables \cite{Knuth1993}.
Each node represents a character of the book.
Each edge indicates that they appear in the same chapter of the book.

The Enron email network represents the email interactions among the staff of Enron Inc \cite{Klimt2004}.
Each node represents an email account. 
Each edge indicates that an email is sent from one account to the other account.  

The jazz network represents the collaborations among jazz musicians \cite{Gleiser2003}.
Each node represents a jazz musician. 
Each edge indicates that two musicians belong to the same band.

The network of network scientists represents the collaborations between researchers in network science \cite{Newman2006}.
Each node represents a researcher.
Two researchers are defined to be adjacent if they have published a co-authored paper cited by one of two popular review papers on network science.
Then, some nodes and edges were added manually by the author of the article ~\cite{Newman2006}.
We only consider the largest connected component of the network. 

The political blog network is the network of blogs on the United States presidential election in 2004  \cite{Adamic2005}.
Each node represents a blog. 
Two blogs are defined to be adjacent if there is at least one hyperlink between the two blogs on their front page.

The airport network consists of nodes representing airports in the world \cite{Openflight.org,ToreOpsahl}.
Two airports are defined to be adjacent if there is a direct commercial flight between the two airports.

The protein network represents the physical interactions among human proteins \cite{Rual2005,Vidal}.
Each node represents a protein. 
Two proteins are defined to be adjacent if they physically interact.  

The Chess network represents the chess matches between players \cite{KONECT}.  
Each node represents a chess player. 
Each edge indicates that they have played at least once. 
 
The Astro-ph network represents the collaborations among the researchers who published a joint paper in the arXiv's astro-ph section  \cite{Leskovec2007}.
Each node represents a researcher. 
Two researchers are defined to be adjacent if they have published a joint paper.

The Internet network represents the network of autonomous systems \cite{KONECT}.
A node represents an autonomous system, which is a group of routers maintained by a network operator. 
Two autonomous systems are defined to be adjacent if they have a logical peering relation.

\section{Results}
We measure the size of a community in two ways: the number of nodes in a community $c$, $n_c$, and the sum of degrees of nodes in a community $c$, $\vol_c$.
In the next two subsections, we consider the $(\qmod_c, n_c)$--test and the $(\qmod_c, \vol_c)$--test.
We show the results for other quality functions in the third subsection.

\subsection{Synthetic networks}
\label{sec:synthetic_networks}
In this section, we examine synthetic networks with planted communities.
We generate networks using the LFR model \cite{Lancichinetti2008}, which places edges such that the node's degree, (i.e., $d_i$), and the number of nodes in a community $c$, (i.e., $n_c$), follow power-law distributions.
We set the power-law exponent for the distributions of $d_i$ and $n_c$ to 2, the average node's degree to 10, the maximum degree to 100 and the range of $n_c$ to $[20, 200]$. 
The networks are composed of $N=10^3$ nodes. 
Each node $i$ has an average fraction $1-\mu$ of neighbours belonging to the same community, where $\mu \in \{0, 0.025, 0.05, \ldots, 1\}$ is a mixing parameter controlling the ``strength'' of community structure. 
With $\mu=0$, all edges are placed within communities, and the community structure is the strongest.
With $\mu=1$, all edges are between different communities. 
We set the extent of overlaps between different communities to zero.

We generate 30 networks using the LFR model at each $\mu$ value.  
For each generated network, we classify the planted communities into significant and insignificant communities by each statistical test. 
Then, we compute the true positive rate (i.e., the fraction of significant communities in the network). 
Finally, we average the true positive rate over the 30 generated networks. 

\begin{figure}
    \centering
    \begin{tabular}{c}
        \begin{minipage}[t][][b]{0.5\hsize}
            \centering
            \includegraphics[width=\hsize]{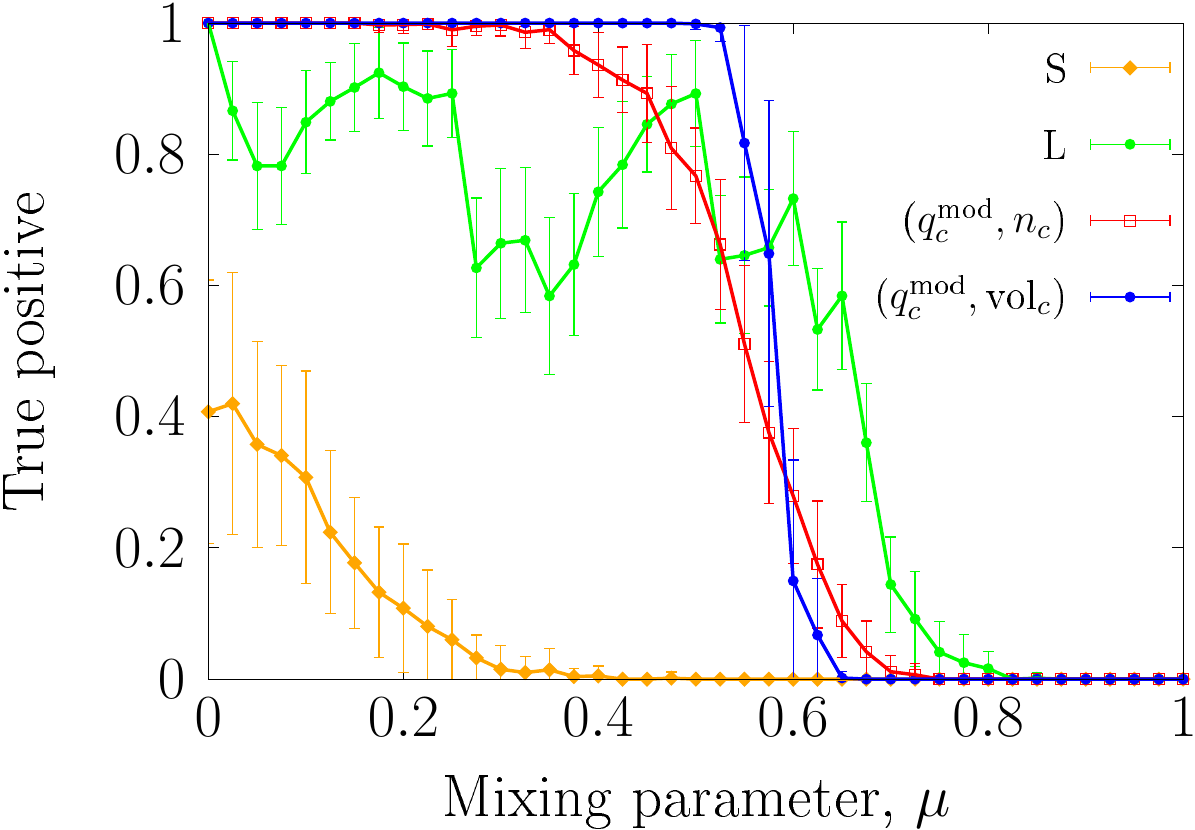}
        \end{minipage}
    \end{tabular}
    \caption{
        True positive rate for the statistical tests applied to the networks generated by the LFR model.
        Legends S, L, $(\qmod_c, n_c)$ and $(\qmod_c, \vol_c)$ indicate the S--test, the L--test, the $(\qmod_c, n_c)$--test and the $(\qmod_c, \vol_c)$--test, respectively. 
        The error bars indicate the $\pm 1$ standard deviation. 
    }
    \label{fig:synthe_tp}
\end{figure}

Figure~\ref{fig:synthe_tp} shows the true positive rate as a function of $\mu$.
The true positive rate for the S--test is smallest for the entire range of $\mu$, indicating that the S--test is the most conservative.
The S--test does not regard all the planted communities as significant even at $\mu=0$ for the following reason.
In the S--test, one detects the strongest community in each randomised network, where the strength of a community is measured by the number of intra-community edges. 
Then, a focal community in the original network is regarded as significant if it is stronger than the majority of the strongest communities detected in the randomised networks.
The strongest communities in the randomised networks often contain almost the largest possible number of intra-community edges, whereas the planted communities do not always even at $\mu = 0$. 
Therefore, the S--test concludes that some planted communities are insignificant.
The true positive rate for the L--test is 1 when $\mu =0$ and ranges between 0.55 and 0.95 for $0 < \mu \leq 0.5$.
The true positive rate for the $(\qmod_c, n_c)$--test and that for the $(\qmod_c, \vol_c)$--test are comparable and close to 1 for $0\leq \mu \leq 0.3$.
In contrast, there is a visible difference between the results for the $(\qmod_c, n_c)$-- and the $(\qmod_c, \vol_c)$--tests for $0.3 < \mu \leq 0.5$. 
This result suggests that the definition of the size of a community may affect the significance of weak communities but not of strong communities.

\subsection{Empirical networks}
\label{sec:empirical_networks}
We apply the statistical tests to the 12 empirical networks listed in Table~\ref{ta:empnet} (see the Data section for details).
In this section, we detect communities by modularity maximisation using the Louvain algorithm \cite{Blondel2008}.
Then, we apply the statistical tests to each detected community.

The fraction of significant communities for each statistical test is shown in Table~\ref{ta:emp_tp}. 
The $(\qmod_c, n_c)$-- and the $(\qmod_c, \vol_c)$--tests identify more significant communities than the S--test and the L--test do in a majority of the 12 empirical networks.
This result indicates that the $(\qmod_c, n_c)$-- and the $(\qmod_c, \vol_c)$--tests are more generous than the S--test and L--test, which is consistent with the results for the LFR model.
This is probably because the ($\qmod_c, n_c$)-- and the ($\qmod_c, \vol_c$)--tests use $\qmod_c$ to evaluate the quality of individual communities, which is consistent with the objective function of modularity maximisation, $\sum_{c=1}^C \qmod_c$.

To quantify the agreement between the $(\qmod_c, n_c)$-- and the $(\qmod_c, \vol_c)$--tests,
we compute the level of agreement defined by $\tau=(C_{11} + C_{00}) / C$, where $C_{00}$ is the number of communities classified as insignificant by both statistical tests and $C_{11}$ is the number of communities classified as significant by both tests.
Note that $0 \leq \tau \leq 1$, $\tau = 1$ if the two tests regard the same set of communities as significant, and $\tau=0$ if the two tests completely disagree.
We compute $\tau$ between each pair of statistical tests for each empirical network and then average $\tau$ over the 12 empirical networks. 
The averaged $\tau$ values are shown in Table~\ref{ta:emp_match}.
We find $\tau=0.42$ between the S--test and the L--test, indicating that the two statistical tests disagree for a majority of communities.
The L--test weakly agrees with the $(\qmod_c, \vol_c)$--test (i.e., $\tau = 0.58$) but disagrees with the other tests for a majority of communities (i.e., $\tau < 0.5$). 
The $\tau$ between the $(\qmod_c, n_c)$-- and the $(\qmod_c, \vol_c)$--tests is large ($\tau=0.84$), suggesting that the significance of a majority of communities is not strongly affected by the definition of the community size.

\subsection{Other quality functions}
\label{sec:other_quality_functions}
In this section, we examine the $(\qint_c, s_c)$--, the $(\qexp_c, s_c)$-- and the $(\qcnd_c, s_c)$--tests, where $s_c$ is either $n_c$ or $\vol_c$.
For the synthetic networks, the true positive rate for the $(\qint_c, n_c)$-- and the $(\qint_c, \vol_c)$--tests is small in the entire range of $\mu$ (Fig.~3).
As is the case for the $S$--test, quality function $\qint_c$ uses the number of intra-community edges.
Some planted communities are regarded as insignificant because randomised networks often contain a community having almost the largest possible number of intra-community edges (Fig.~\ref{fig:lfr}(c)).
The quality function $\qexp_c$ is the largest when the community $c$ is disconnected from the other nodes.
Randomised networks often contain many disconnected components, yielding a large value of $\qexp_c$ (Fig.~\ref{fig:lfr}(d)).
Therefore, the true positive rate for the $(\qexp_c, n_c)$-- and the $(\qexp_c, \vol_c)$--tests is also close to zero in the entire range of $\mu$.
In contrast to $(\qint_c, s_c)$-- and $(\qexp_c, s_c)$--tests, the $(\qcnd_c, n_c)$-- and $(\qcnd_c, \vol_c)$--tests yield the true positive rate close to one when $\mu \leq 0.3$.
These results suggest that the results considerably depend on the quality function. 
For all the $(q,s)$--tests, the definition of community size (i.e., $n_c$ or $\vol_c$) does not strongly influence the true positive rate.

\begin{figure}
    \centering
    \begin{tabular}{c}
        \begin{minipage}[t][][b]{0.5\hsize}
            \centering
            \includegraphics[width=\hsize]{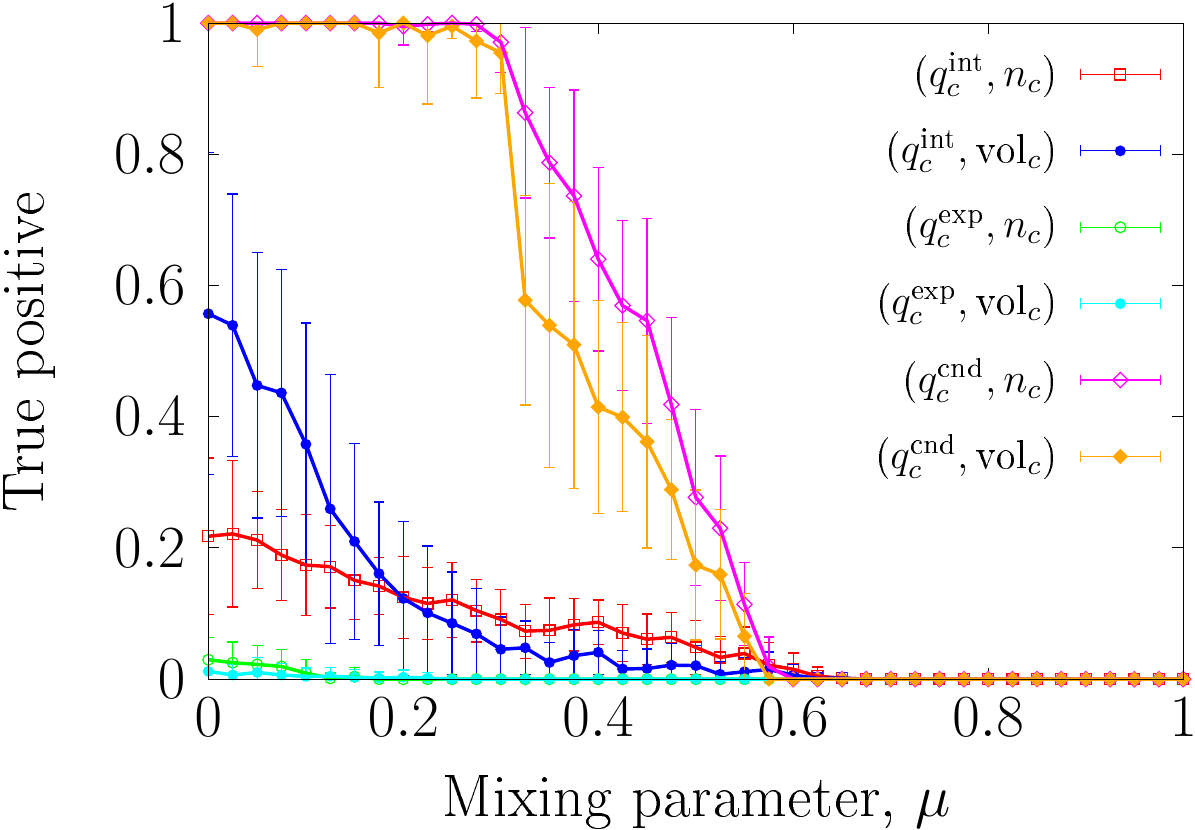}
        \end{minipage}
    \end{tabular}
    \caption{
        True positive rate as a function of mixing parameter, $\mu$, for 
        the six $(q,s)$--tests. 
    }
    \label{fig:synthe_tp_other_qfunc}
\end{figure}

For the empirical networks, we first detect communities by maximising $q$, where $q$ is either $\qint_c$, $\qexp_c$ or $\qcnd_c$, using the variant of the Kernighan--Lin algorithm (see the Other statistical test sections).
Then, we apply the $(q,s)$--test to each detected community.
The results for the $(\qint_c, s_c)$--, the $(\qexp_c, s_c)$-- and the $(\qcnd_c, s_c)$--tests applied to the 12 empirical networks are shown in Table~\ref{ta:emp_tp}.
For all the networks, the $(\qcnd_c, s_c)$--test regards more communities as significant than the $(\qexp_c, s_c)$-- and the $(\qcnd_c, s_c)$--tests, where $s_c$ is either $n_c$ or $\vol_c$. 
This result is consistent with those obtained for the synthetic networks (Fig.~\ref{fig:synthe_tp_other_qfunc}).
For each quality function $q$, the level of agreement (i.e., $\tau$) between the different definitions of the community size (i.e., $n_c$ or $\vol_c$) is shown in Table~\ref{ta:emp_match_other_qfunc}.
For most empirical networks, the agreement $\tau$ is larger than $0.8$, indicating that the results of the statistical test do not strongly depend on the definition of community size in most cases.

\section{Discussion}
\label{sec:discussion}
We proposed a non-parametric statistical test, called the $(q,s)$--test, for the significance of individual communities, which accounts for the correlation between the quality and the size of single communities.
We demonstrated our test with several quality functions $q$ including the one defined as the contribution of a single community to the modularity.
In fact, the $(q,s)$--test accepts different quality functions for individual communities such as those described in the previous literature ~\cite{Leskovec2010,Chen2014,Lambiotte2014,Zhang2014,Yang2015}. 
In addition, the $(q,s)$--test does not demand how communities should be detected in a given network. 
We note that $q$ that is consistent with the objective function for community detection should be used because the former is maximised in the $(q, s)$--test and the latter is maximised in community detection.

We have used two definitions of the size of a community, i.e., the number of nodes in a community (i.e., $n_c$), and the sum of degrees of nodes in a community (i.e., $\vol_c$). 
For degree-homogeneous networks, the choice does not matter because $n_c \propto \vol_c$.  
However, for degree-heterogeneous networks, significant communities may considerably depend on whether we use $n_c$ or $\vol_c$. 
If $q$ explicitly uses its own measure of the size of a community, we should probably adopt the corresponding definition of the community size in the $(q, s)$--test. 
If a measure of community size is not explicit, we suggest that one selects a measure of community size that is more strongly correlated with $q$ than others.
If $q$ is correlated with multiple quantities (e.g. both $n_c$ and $\vol_c$) that are not perfectly correlated with each other, one can extend the $(q,s)$--test by adopting multivariate Gaussian kernels with three or more variables instead of bivariate Gaussian kernels.
A downside of this approach is that we would need more data to reliably estimate the distribution of $(q,s)$, where $s$ is at least two-dimensional.

We can adopt the $(q,s)$--test to assess the significance of other structures of networks, such as bipartite communities \cite{Newman2007} and core-periphery structure \cite{Borgatti2000,Rombach2017,Kojaku2017b}, provided that the quality function for the individual structure (e.g., a single bipartite community) is explicitly defined.
In fact, we applied a variant of the $(q, s)$--test to core-periphery structure in our previous study \cite{Kojaku2017b}.

Robustness of community structure against random perturbations (e.g., addition, removal and rewiring of edges) is an alternative measure of the significance of communities \cite{Gfeller2005,Karrer2008,Yang2015}.
With this approach, if small perturbations do not considerably change communities, then the communities are regarded as significant.
Statistical tests based on quality functions including the $(q, s)$--test and those based on robustness may provide different results \cite{Karrer2008}.
As is the case of quality functions, the robustness of an individual community may be correlated with the size of a community.
For example, removal of a small number of intra-community edges may destroy small communities, whereas large communities may survive the removal of more intra-community edges. 
If this is the case, it may be worthwhile to inform a robustness--based test of individual communities by the dependence of the robustness measure on the size of a community.

\section*{Competing interests}
The authors declare no competing interests.

\section*{Author contributions}
N.~M. conceived and designed the research; S.~K. performed the computational experiments; N.~M. and S.~K. wrote the paper.

\section*{Acknowledgement}
N.~M. acknowledges the support provided through JST, CREST, and JST, ERATO, Kawarabayashi Large Graph Project.


\begin{thebibliography}{10}
\expandafter\ifx\csname url\endcsname\relax
  \def\url#1{\texttt{#1}}\fi
\expandafter\ifx\csname urlprefix\endcsname\relax\def\urlprefix{URL }\fi
\providecommand{\bibinfo}[2]{#2}
\providecommand{\eprint}[2][]{\url{#2}}

\bibitem{Newman2010}
\bibinfo{author}{Newman, M. E.~J.}
\newblock \emph{\bibinfo{title}{{Networks: An Introduction}}}
  (\bibinfo{publisher}{Oxford University Press}, \bibinfo{address}{Oxford},
  \bibinfo{year}{2010}).

\bibitem{Barabasi2016}
\bibinfo{author}{Barab{\'a}si, A.~L.}
\newblock \emph{\bibinfo{title}{Network Science}}
  (\bibinfo{publisher}{Cambridge University Press},
  \bibinfo{address}{Cambridge}, \bibinfo{year}{2016}).

\bibitem{Fortunato2010}
\bibinfo{author}{Fortunato, S.}
\newblock \bibinfo{journal}{\bibinfo{title}{{Community detection in graphs}}}.
\newblock {\emph{\JournalTitle{Phys. Rep.}}} \textbf{\bibinfo{volume}{486}},
  \bibinfo{pages}{75--174} (\bibinfo{year}{2010}).

\bibitem{Fortunato2016}
\bibinfo{author}{Fortunato, S.} \& \bibinfo{author}{Hric, D.}
\newblock \bibinfo{journal}{\bibinfo{title}{{Community detection in networks: A
  user guide}}}.
\newblock {\emph{\JournalTitle{Phys. Rep.}}} \textbf{\bibinfo{volume}{659}},
  \bibinfo{pages}{1--44} (\bibinfo{year}{2016}).

\bibitem{Jonsson2006}
\bibinfo{author}{Jonsson, P.~F.}, \bibinfo{author}{Cavanna, T.},
  \bibinfo{author}{Zicha, D.} \& \bibinfo{author}{Bates, P.~A.}
\newblock \bibinfo{journal}{\bibinfo{title}{{Cluster analysis of networks
  generated through homology: automatic identification of important protein
  communities involved in cancer metastasis}}}.
\newblock {\emph{\JournalTitle{BMC Bioinf.}}} \textbf{\bibinfo{volume}{7}},
  \bibinfo{pages}{2} (\bibinfo{year}{2006}).

\bibitem{Guimera2005}
\bibinfo{author}{Guimer{\`{a}}, R.}, \bibinfo{author}{Mossa, S.},
  \bibinfo{author}{Turtschi, A.} \& \bibinfo{author}{Amaral, L. A.~N.}
\newblock \bibinfo{journal}{\bibinfo{title}{The worldwide air transportation
  network: anomalous centrality, community structure, and cities' global
  roles}}.
\newblock {\emph{\JournalTitle{Proc. Natl. Acad. Sci. USA}}}
  \textbf{\bibinfo{volume}{102}}, \bibinfo{pages}{7794--7799}
  (\bibinfo{year}{2005}).

\bibitem{Newman2006}
\bibinfo{author}{Newman, M. E.~J.}
\newblock \bibinfo{journal}{\bibinfo{title}{{Finding community structure in
  networks using the eigenvectors of matrices}}}.
\newblock {\emph{\JournalTitle{Phys. Rev. E}}} \textbf{\bibinfo{volume}{74}},
  \bibinfo{pages}{036104} (\bibinfo{year}{2006}).

\bibitem{Lancichinetti2010}
\bibinfo{author}{Lancichinetti, A.}, \bibinfo{author}{Radicchi, F.} \&
  \bibinfo{author}{Ramasco, J.~J.}
\newblock \bibinfo{journal}{\bibinfo{title}{{Statistical significance of
  communities in networks}}}.
\newblock {\emph{\JournalTitle{Phys. Rev. E}}} \textbf{\bibinfo{volume}{81}},
  \bibinfo{pages}{046110} (\bibinfo{year}{2010}).

\bibitem{Lancichinetti2011}
\bibinfo{author}{Lancichinetti, A.}, \bibinfo{author}{Radicchi, F.},
  \bibinfo{author}{Ramasco, J.~J.} \& \bibinfo{author}{Fortunato, S.}
\newblock \bibinfo{journal}{\bibinfo{title}{{Finding statistically significant
  communities in networks}}}.
\newblock {\emph{\JournalTitle{PLOS ONE}}} \textbf{\bibinfo{volume}{6}},
  \bibinfo{pages}{e18961} (\bibinfo{year}{2011}).

\bibitem{Spirin2003}
\bibinfo{author}{Spirin, V.} \& \bibinfo{author}{Mirny, L.~A.}
\newblock \bibinfo{journal}{\bibinfo{title}{{Protein complexes and functional
  modules in molecular networks}}}.
\newblock {\emph{\JournalTitle{Proc. Natl. Acad. Sci. USA}}}
  \textbf{\bibinfo{volume}{100}}, \bibinfo{pages}{12123--12128}
  (\bibinfo{year}{2003}).

\bibitem{Wang2008}
\bibinfo{author}{Wang, B.} \emph{et~al.}
\newblock \bibinfo{title}{{Spatial scan statistics for graph clustering}}.
\newblock In \emph{\bibinfo{booktitle}{Proc. 2008 SIAM Int. Conf. Data
  Mining}}, \bibinfo{pages}{727--738} (\bibinfo{publisher}{SIAM},
  \bibinfo{address}{Philadelphia}, \bibinfo{year}{2008}).

\bibitem{Zhao2011}
\bibinfo{author}{Zhao, Y.}, \bibinfo{author}{Levina, E.} \&
  \bibinfo{author}{Zhu, J.}
\newblock \bibinfo{journal}{\bibinfo{title}{{Community extraction for social
  networks.}}}
\newblock {\emph{\JournalTitle{Proc. Natl. Acad. Sci. USA}}}
  \textbf{\bibinfo{volume}{108}}, \bibinfo{pages}{7321--7326}
  (\bibinfo{year}{2011}).

\bibitem{Leskovec2010}
\bibinfo{author}{Leskovec, J.}, \bibinfo{author}{Lang, K.~J.} \&
  \bibinfo{author}{Mahoney, M.~W.}
\newblock \bibinfo{title}{{Empirical comparison of algorithms for network
  community detection}}.
\newblock In \emph{\bibinfo{booktitle}{Proc. 19th Int. Conf. World Wide Web}},
  \bibinfo{pages}{631--640} (\bibinfo{publisher}{ACM, New York},
  \bibinfo{year}{2010}).

\bibitem{Yang2015}
\bibinfo{author}{Yang, J.} \& \bibinfo{author}{Leskovec, J.}
\newblock \bibinfo{journal}{\bibinfo{title}{{Defining and evaluating network
  communities based on ground-truth}}}.
\newblock {\emph{\JournalTitle{Know. Inf. Syst.}}}
  \textbf{\bibinfo{volume}{42}}, \bibinfo{pages}{181--213}
  (\bibinfo{year}{2015}).

\bibitem{Lancichinetti2008}
\bibinfo{author}{Lancichinetti, A.}, \bibinfo{author}{Fortunato, S.} \&
  \bibinfo{author}{Radicchi, F.}
\newblock \bibinfo{journal}{\bibinfo{title}{{Benchmark graphs for testing
  community detection algorithms}}}.
\newblock {\emph{\JournalTitle{Phys. Rev. E}}} \textbf{\bibinfo{volume}{78}},
  \bibinfo{pages}{046110} (\bibinfo{year}{2008}).

\bibitem{Wand1993}
\bibinfo{author}{Wand, M.~P.} \& \bibinfo{author}{Jones, M.~C.}
\newblock \bibinfo{journal}{\bibinfo{title}{{Comparison of smoothing
  parameterizations in bivariate kernel density estimation}}}.
\newblock {\emph{\JournalTitle{J. Am. Stat. Assoc.}}}
  \textbf{\bibinfo{volume}{88}}, \bibinfo{pages}{520--528}
  (\bibinfo{year}{1993}).

\bibitem{Parzen1962}
\bibinfo{author}{Parzen, E.}
\newblock \bibinfo{journal}{\bibinfo{title}{{On estimation of a probability
  density function and mode}}}.
\newblock {\emph{\JournalTitle{Annal. Math. Stat.}}}
  \textbf{\bibinfo{volume}{33}}, \bibinfo{pages}{1065--1076}
  (\bibinfo{year}{1962}).

\bibitem{Park1990}
\bibinfo{author}{Park, B.~U.} \& \bibinfo{author}{Marron, J.~S.}
\newblock \bibinfo{journal}{\bibinfo{title}{{Comparison of data-driven
  bandwidth selectors}}}.
\newblock {\emph{\JournalTitle{J. Am. Stat. Assoc.}}}
  \textbf{\bibinfo{volume}{85}}, \bibinfo{pages}{66--72}
  (\bibinfo{year}{1990}).

\bibitem{Jones1996}
\bibinfo{author}{Jones, M.~C.}, \bibinfo{author}{Marron, J.~S.} \&
  \bibinfo{author}{Sheather, S.~J.}
\newblock \bibinfo{journal}{\bibinfo{title}{{A brief survey of bandwidth
  selection for density estimation}}}.
\newblock {\emph{\JournalTitle{J. Am. Stat. Assoc.}}}
  \textbf{\bibinfo{volume}{91}}, \bibinfo{pages}{401--407}
  (\bibinfo{year}{1996}).

\bibitem{Scott2012}
\bibinfo{author}{Scott, D.~W.}
\newblock \emph{\bibinfo{title}{{Multivariate density estimation and
  visualization}}} (\bibinfo{publisher}{Springer}, \bibinfo{address}{Berlin},
  \bibinfo{year}{2012}).

\bibitem{Sidak1967}
\bibinfo{author}{{\v{S}}id{\'{a}}k, Z.}
\newblock \bibinfo{journal}{\bibinfo{title}{{Rectangular confidence regions for
  the means of multivariate normal distributions}}}.
\newblock {\emph{\JournalTitle{J. Am. Stat. Assoc.}}}
  \textbf{\bibinfo{volume}{62}}, \bibinfo{pages}{626--633}
  (\bibinfo{year}{1967}).

\bibitem{Miller2011}
\bibinfo{author}{Miller, J.~C.} \& \bibinfo{author}{Hagberg, A.}
\newblock \bibinfo{title}{{Efficient generation of networks with given expected
  degrees}}.
\newblock In \bibinfo{editor}{Frieze, A.}, \bibinfo{editor}{Horn, P.} \&
  \bibinfo{editor}{Pra{\l}at, P.} (eds.) \emph{\bibinfo{booktitle}{Algorithms
  and Models for the Web Graph}}, vol. \bibinfo{volume}{6732 LNCS},
  \bibinfo{pages}{115--126} (\bibinfo{publisher}{Springer Berlin Heidelberg},
  \bibinfo{address}{Berlin, Heidelberg}, \bibinfo{year}{2011}).

\bibitem{Networkit}
\bibinfo{author}{Staudt, C.~L.}, \bibinfo{author}{Sazonovs, A.} \&
  \bibinfo{author}{Meyerhenke, H.}
\newblock \bibinfo{journal}{\bibinfo{title}{Networkit: A tool suite for
  large-scale complex network analysis}}.
\newblock {\emph{\JournalTitle{Network Science}}} \textbf{\bibinfo{volume}{4}},
  \bibinfo{pages}{508--530} (\bibinfo{year}{2016}).

\bibitem{Networkx}
\bibinfo{author}{Hagberg, A.~A.}, \bibinfo{author}{Schult, D.~A.} \&
  \bibinfo{author}{Swart, P.~J.}
\newblock \bibinfo{title}{Exploring network structure, dynamics, and function
  using networkx}.
\newblock In \bibinfo{editor}{Varoquaux, G.}, \bibinfo{editor}{Vaught, T.} \&
  \bibinfo{editor}{Millman, J.} (eds.) \emph{\bibinfo{booktitle}{Proc. 7th
  Python in Sci. Conf.}}, \bibinfo{pages}{11 -- 15}
  (\bibinfo{address}{Pasadena, CA USA}, \bibinfo{year}{2008}).

\bibitem{KONECT}
\bibinfo{note}{J. Kunegis. Available at http://konect.uni-koblenz.de [Accessed:
  2 Sep 2017]}.

\bibitem{Blondel2008}
\bibinfo{author}{Blondel, V.~D.}, \bibinfo{author}{Guillaume, J.-L.},
  \bibinfo{author}{Lambiotte, R.} \& \bibinfo{author}{Lefebvre, E.}
\newblock \bibinfo{journal}{\bibinfo{title}{{Fast unfolding of communities in
  large networks}}}.
\newblock {\emph{\JournalTitle{J. Stat. Mech.}}}
  \textbf{\bibinfo{volume}{2008}}, \bibinfo{pages}{P10008}
  (\bibinfo{year}{2008}).

\bibitem{Karrer2011}
\bibinfo{author}{Karrer, B.} \& \bibinfo{author}{Newman, M. E.~J.}
\newblock \bibinfo{journal}{\bibinfo{title}{Stochastic blockmodels and
  community structure in networks}}.
\newblock {\emph{\JournalTitle{Phys. Rev. E}}} \textbf{\bibinfo{volume}{83}},
  \bibinfo{pages}{016107} (\bibinfo{year}{2011}).

\bibitem{Luxburg2007}
\bibinfo{author}{von Luxburg, U.}
\newblock \bibinfo{journal}{\bibinfo{title}{{A tutorial on spectral
  clustering}}}.
\newblock {\emph{\JournalTitle{Stat. Comput.}}} \textbf{\bibinfo{volume}{17}},
  \bibinfo{pages}{395--416} (\bibinfo{year}{2007}).

\bibitem{Kernighan1970}
\bibinfo{author}{Kernighan, B.~W.} \& \bibinfo{author}{Lin, S.}
\newblock \bibinfo{journal}{\bibinfo{title}{An efficient heuristic procedure
  for partitioning graphs}}.
\newblock {\emph{\JournalTitle{Bell Syst. Tech. J.}}}
  \textbf{\bibinfo{volume}{49}}, \bibinfo{pages}{291--307}
  (\bibinfo{year}{1970}).

\bibitem{Zachary1977}
\bibinfo{author}{Zachary, W.~W.}
\newblock \bibinfo{journal}{\bibinfo{title}{{An information flow model for
  conflict and fission in small groups}}}.
\newblock {\emph{\JournalTitle{J. Anthropol. Res.}}}
  \textbf{\bibinfo{volume}{33}}, \bibinfo{pages}{452--473}
  (\bibinfo{year}{1977}).

\bibitem{Lusseau2003}
\bibinfo{author}{Lusseau, D.} \emph{et~al.}
\newblock \bibinfo{journal}{\bibinfo{title}{{The bottlenose dolphin community
  of doubtful sound features a large proportion of long-lasting
  associations}}}.
\newblock {\emph{\JournalTitle{Behav. Ecol. Sociobiol.}}}
  \textbf{\bibinfo{volume}{54}}, \bibinfo{pages}{396--405}
  (\bibinfo{year}{2003}).

\bibitem{Knuth1993}
\bibinfo{author}{Knuth, D.~E.}
\newblock \emph{\bibinfo{title}{{The Stanford GraphBase: A Platform for
  Combinatorial Computing}}} (\bibinfo{publisher}{ACM Press, New York},
  \bibinfo{year}{1993}).

\bibitem{Klimt2004}
\bibinfo{author}{Klimt, B.} \& \bibinfo{author}{Yang, Y.}
\newblock \bibinfo{title}{{The Enron corpus: A new dataset for email
  classification research}}.
\newblock In \emph{\bibinfo{booktitle}{Proc. 15th European Conf. Machine
  Learning}}, \bibinfo{pages}{217--226} (\bibinfo{publisher}{Springer, Berlin},
  \bibinfo{year}{2004}).

\bibitem{Gleiser2003}
\bibinfo{author}{Gleiser, P.~M.} \& \bibinfo{author}{Danon, L.}
\newblock \bibinfo{journal}{\bibinfo{title}{{Community structure in jazz}}}.
\newblock {\emph{\JournalTitle{Adv. Comp. Syst.}}}
  \textbf{\bibinfo{volume}{6}}, \bibinfo{pages}{565--573}
  (\bibinfo{year}{2003}).

\bibitem{Adamic2005}
\bibinfo{author}{Adamic, L.~A.} \& \bibinfo{author}{Glance, N.}
\newblock \bibinfo{title}{The political blogosphere and the 2004 u.s. election:
  divided they blog}.
\newblock In \emph{\bibinfo{booktitle}{Proc. 3rd Int. Workshop on Link
  Discovery}}, \bibinfo{pages}{36--43} (\bibinfo{publisher}{ACM, New York},
  \bibinfo{year}{2005}).

\bibitem{Openflight.org}
\bibinfo{note}{J. Patokallio. Available at http://openflights.org [Accessed: 24
  Sep 2016]}.

\bibitem{ToreOpsahl}
\bibinfo{note}{{T. Opsahl. Available at
  https://toreopsahl.com/2011/08/12/why-anchorage-is-not-that-important-binary-ties-and-sample-selection
  [Accessed: 24 Sep 2016]}}.

\bibitem{Rual2005}
\bibinfo{author}{Rual, J.} \emph{et~al.}
\newblock \bibinfo{journal}{\bibinfo{title}{Towards a proteome-scale map of the
  human protein-protein interaction network}}.
\newblock {\emph{\JournalTitle{Nature}}} \textbf{\bibinfo{volume}{437}},
  \bibinfo{pages}{1173--1178} (\bibinfo{year}{2005}).

\bibitem{Vidal}
\bibinfo{note}{A. Ma'ayan. Available at
  http://research.mssm.edu/maayan/datasets/qualitative\_networks.shtml
  [Accessed: 2 Sep 2017]}.

\bibitem{Leskovec2007}
\bibinfo{author}{Leskovec, J.}, \bibinfo{author}{Kleinberg, J.} \&
  \bibinfo{author}{Faloutsos, C.}
\newblock \bibinfo{journal}{\bibinfo{title}{{Graph evolution: densification and
  shrinking diameters}}}.
\newblock {\emph{\JournalTitle{ACM Trans. Knowl. Discov. Data}}}
  \textbf{\bibinfo{volume}{1}}, \bibinfo{pages}{2} (\bibinfo{year}{2007}).

\bibitem{Chen2014}
\bibinfo{author}{Chen, M.}, \bibinfo{author}{Kuzmin, K.} \&
  \bibinfo{author}{Szymanski, B.~K.}
\newblock \bibinfo{journal}{\bibinfo{title}{{Community detection via
  maximization of modularity and its variants}}}.
\newblock {\emph{\JournalTitle{IEEE Trans. Comput. Soc. Syst.}}}
  \textbf{\bibinfo{volume}{1}}, \bibinfo{pages}{46--65} (\bibinfo{year}{2014}).

\bibitem{Lambiotte2014}
\bibinfo{author}{Lambiotte, R.}, \bibinfo{author}{Delvenne, J.~C.} \&
  \bibinfo{author}{Barahona, M.}
\newblock \bibinfo{journal}{\bibinfo{title}{{Random walks, markov processes and
  the multiscale modular organization of complex networks}}}.
\newblock {\emph{\JournalTitle{IEEE Trans. Netw. Sci. Eng.}}}
  \textbf{\bibinfo{volume}{1}}, \bibinfo{pages}{76--90} (\bibinfo{year}{2014}).

\bibitem{Zhang2014}
\bibinfo{author}{Zhang, P.} \& \bibinfo{author}{Moore, C.}
\newblock \bibinfo{journal}{\bibinfo{title}{{Scalable detection of
  statistically significant communities and hierarchies, using message passing
  for modularity}}}.
\newblock {\emph{\JournalTitle{Proc. Natl. Acad. Sci. USA}}}
  \textbf{\bibinfo{volume}{111}}, \bibinfo{pages}{18144--18149}
  (\bibinfo{year}{2014}).

\bibitem{Newman2007}
\bibinfo{author}{Newman, M. E.~J.} \& \bibinfo{author}{Leicht, E.~A.}
\newblock \bibinfo{journal}{\bibinfo{title}{{Mixture models and exploratory
  analysis in networks}}}.
\newblock {\emph{\JournalTitle{Proc. Natl. Acad. Sci. USA}}}
  \textbf{\bibinfo{volume}{104}}, \bibinfo{pages}{9564--9569}
  (\bibinfo{year}{2007}).

\bibitem{Borgatti2000}
\bibinfo{author}{Borgatti, S.~P.} \& \bibinfo{author}{Everett, M.~G.}
\newblock \bibinfo{journal}{\bibinfo{title}{{Models of core/periphery
  structures}}}.
\newblock {\emph{\JournalTitle{Soc. Netw.}}} \textbf{\bibinfo{volume}{21}},
  \bibinfo{pages}{375--395} (\bibinfo{year}{2000}).

\bibitem{Rombach2017}
\bibinfo{author}{Rombach, M.~P.}, \bibinfo{author}{Porter, M.~A.},
  \bibinfo{author}{Fowler, J.~H.} \& \bibinfo{author}{Mucha, P.~J.}
\newblock \bibinfo{journal}{\bibinfo{title}{Core-periphery structure in
  networks (revisited)}}.
\newblock {\emph{\JournalTitle{SIAM Rev.}}} \textbf{\bibinfo{volume}{59}},
  \bibinfo{pages}{619--646} (\bibinfo{year}{2017}).

\bibitem{Kojaku2017b}
\bibinfo{author}{Kojaku, S.} \& \bibinfo{author}{Masuda, N.}
\newblock \bibinfo{journal}{\bibinfo{title}{{Core-periphery structure requires
  something else in the network}}}.
\newblock {\emph{\JournalTitle{arXiv}}} \bibinfo{note}{Preprint
  arXiv:1710.07076 (2017)}.

\bibitem{Gfeller2005}
\bibinfo{author}{Gfeller, D.}, \bibinfo{author}{Chappelier, J.~C.} \&
  \bibinfo{author}{{De Los Rios}, P.}
\newblock \bibinfo{journal}{\bibinfo{title}{{Finding instabilities in the
  community structure of complex networks}}}.
\newblock {\emph{\JournalTitle{Phys. Rev. E}}} \textbf{\bibinfo{volume}{72}},
  \bibinfo{pages}{056135} (\bibinfo{year}{2005}).

\bibitem{Karrer2008}
\bibinfo{author}{Karrer, B.}, \bibinfo{author}{Levina, E.} \&
  \bibinfo{author}{Newman, M. E.~J.}
\newblock \bibinfo{journal}{\bibinfo{title}{Robustness of community structure
  in networks}}.
\newblock {\emph{\JournalTitle{Phys. Rev. E}}} \textbf{\bibinfo{volume}{77}},
  \bibinfo{pages}{046119} (\bibinfo{year}{2008}).

\end{thebibliography}

\clearpage
\clearpage
\clearpage
\clearpage
\begin{table}
\centering
\caption{
    Properties of 12 empirical networks.
    Column $C$ indicates the number of communities detected by the Louvain algorithm. 
    Columns $n_c$ and $\vol_c$ indicate the number of nodes in a community and the sum of degrees of nodes in a community, respectively.  
}
\label{ta:empnet}
\scalebox{0.9}{
\begin{tabular}{l|cccccccc}
\multirow{2}{4em}{\centering Network} & \multirow{2}{4em}{\centering $N$} & \multirow{2}{4em}{\centering $M$} & \multirow{2}{4em}{\centering $C$} & \multicolumn{2}{c}{$n_c$} &  & \multicolumn{2}{c}{$\vol_c$} \\ \cline{5-6} \cline{8-9}
& & & & Min & Max && Min & Max\\ \hline \hline
Karate  \cite{Zachary1977}                 & 34         & 78         & 3     & 5     & 17         & & 16 & 78\\
Dolphin  \cite{Lusseau2003}                & 62         & 159         & 4     & 7     & 22         & & 37 & 123\\
Les Mis\'{e}rables  \cite{Knuth1993}             & 77         & 254         & 10     & 2     & 16         & & 3 & 147\\
Email \cite{Klimt2004}                     & 151         & 1527         & 6     & 16     & 50         & & 258 & 1081\\
Jazz \cite{Gleiser2003}                 & 198         & 2742         & 6     & 3     & 63         & & 9 & 2029\\
Network science \cite{Newman2006}     & 379         & 914         & 11     & 6     & 65         & & 27 & 290\\
Blog \cite{Adamic2005}                     & 1222         & 16{,}714     & 2     & 565     & 657         & & 15,755 & 17,673\\
Airport \cite{Openflight.org,ToreOpsahl}         & 2939         & 15{,}677     & 20     & 2     & 712         & & 2 & 12,638\\
Protein \cite{Rual2005,Vidal}                 & 3023         & 6149         & 161     & 2     & 312         & & 2 & 1832\\
Chess \cite{KONECT}                     & 7115         & 55{,}779     & 409     & 2     & 812         & & 3 & 23,034\\
Astro-ph (co-authorship) \cite{Leskovec2007}         & 18{,}771     & 198{,}050     & 116     & 2     & 3547         & & 2 & 98,628\\
Internet \cite{KONECT}                     & 34{,}761     & 107{,}720     & 65     & 4     & 13{,}710     & & 7 & 106,881\\ \hline 
\end{tabular}
}
\end{table}
\clearpage
\begin{table}
\centering
\caption{
Fraction of significant communities identified by the S--test, the L--test, the $(\qmod,s)$--test, the $(\qint,s)$--test,  the $(\qexp,s)$--test and the $(\qcnd,s)$--test in the 12 empirical networks.
The hyphen indicates that the test did not terminate within 64 days on our computer (Intel 2.6GHz Sandy Bridge processors and 4GB of memory).
}
\label{ta:emp_tp}
\setlength{\tabcolsep}{5pt}
\centering
\begin{tabular}{l|cccccccccccccccc}
\multirow{2}{*}{Network} & \multirow{2}{*}{\centering S} && \multirow{2}{*}{\centering L} && \multicolumn{2}{c}{$\qmod_c$}  && \multicolumn{2}{c}{$\qint_c$}  & & \multicolumn{2}{c}{$\qexp_c$} & & \multicolumn{2}{c}{$\qcnd_c$} \\ \cline{6-7} \cline{9-10} \cline{12-13} \cline{15-16}
            &&& & & $n_c$ & $\vol_c$ & & $n_c$ & $\vol_c$  & & $n_c$ & $\vol_c$  & & $n_c$ & $\vol_c$  \\ \hline \hline
Karate              & 1.00 && 0.33 && 0.67 & 1.00 && 0.00 & 0.00 && 0.00 & 0.00 && 0.00 & 0.33\\
Dolphin          & 1.00 && 0.50 && 1.00 & 0.75 && 0.00 & 0.00 && 0.00 & 0.00 && 0.50 & 0.50\\
Les Mis\'{e}rables      & 0.40 && 0.40 && 0.40 & 0.60 && 0.20 & 0.40 && 0.00 & 0.00 && 0.50 & 0.40\\
Enron              & 1.00 && 0.00 && 1.00 & 1.00 && 0.33 & 0.67 && 0.00 & 0.00 && 1.00 & 1.00\\
Jazz              & 0.67 && 0.67 && 0.67 & 1.00 && 0.67 & 0.83 && 0.00 & 0.00 && 1.00 & 1.00\\
Netscience          & 1.00 && 0.64 && 1.00 & 1.00 && 0.91 & 0.82 && 0.09 & 0.09 && 0.91 & 1.00\\
Blog              & 0.00 && 1.00 && 1.00 & 1.00 && 0.50 & 0.50 && 0.00 & 0.00 && 1.00 & 1.00\\
Airport          & 0.00 && 0.60 && 0.70 & 0.80 && 0.15 & 0.55 && 0.00 & 0.00 && 0.40 & 0.20\\
Protein          & 0.00 && 0.35 && 0.14 & 0.22 && 0.03 & 0.12 && 0.01 & 0.01 && 0.00 & 0.00\\
Chess              & 0.00 && 0.25 && 0.13 & 0.15 && 0.36 & 0.58 && 0.00 & 0.00 && 0.01 & 0.03\\
Astro-ph          & --   && 0.61 && 0.24 & 0.53 && 1.00 & 1.00 && 0.00 & 0.00 && 0.33 & 0.12\\
Internet          & --   && 0.55 && 0.65 & 0.60 && 0.00 & 0.18 && 0.00 & 0.00 && 0.00 & 0.02\\ \hline
\end{tabular}
\end{table}
\clearpage
\begin{table}
\centering
\caption{
Agreement between pairs of statistical tests. 
}
\label{ta:emp_match}
\setlength{\tabcolsep}{10pt}
\begin{tabular}{c|cccc}
 Test    &     \multirow{1}{4em}{\centering S}     & \multirow{1}{4em}{\centering L} & \multirow{1}{4em}{\centering $(\qmod_c, n_c)$} & \multirow{1}{4em}{\centering $(\qmod_c, \vol_c)$} \\ \hline \hline
S         & 1.00 & 0.42 & 0.73 & 0.66\\
L         & 0.42 & 1.00 & 0.49 & 0.58\\
($\qmod_c, n_c$)    & 0.73 & 0.49 & 1.00 & 0.84\\
($\qmod_c, \vol_c$)    & 0.66 & 0.58 & 0.84 & 1.00\\ \hline
\end{tabular}
\end{table}
\clearpage
\begin{table}
\centering
\caption{
Agreement between the $(q_c,n_c)$--test and the $(q_c,\vol_c)$--test. 
}
\label{ta:emp_match_other_qfunc}
\setlength{\tabcolsep}{7pt}
\begin{tabular}{l|ccc}
\multirow{2}{*}{Network} & \multicolumn{3}{c}{$q_c$} \\\cline{2-4} 
         & $\qint_c$ & $\qexp_c$ & $\qcnd_c$\\ \hline \hline
Karate             & 1.00 & 1.00 & 0.67\\
Dolphin         & 1.00 & 1.00 & 1.00\\
Les Mis\'{e}rables     & 0.60 & 1.00 & 0.90\\
Enron             & 0.67 & 1.00 & 1.00\\
Jazz             & 0.50 & 1.00 & 1.00\\
Netscience         & 0.73 & 1.00 & 0.91\\
Blog             & 1.00 & 1.00 & 1.00\\
Airport         & 0.60 & 1.00 & 0.60\\
Protein         & 0.90 & 0.99 & 1.00\\
Chess             & 0.77 & 1.00 & 0.98\\
Astro-ph         & 1.00 & 1.00 & 0.76\\
Internet         & 0.82 & 1.00 & 0.98\\ \hline
\end{tabular}
\end{table}

\clearpage
 


\section*{Supplementary material}

# Supplementary Information: A generalised significance test for individual communities in networks

Sadamori Kojaku and Naoki Masuda

## I. $p$-VALUE UNDER THE NULL MODEL

In our statistical test, the $p$-value is given by

$$F_{\tilde{q}}(q_c) = \int_{q_c}^{\infty} P(\tilde{q}|s_c) \mathrm{d}\tilde{q}, \tag{1}$$

where $q_c$ is the quality of a focal community $c$, $\tilde{q}$ is the quality of a community detected in the randomised networks and $s_c$ is the size of the focal community. Function $F_{\tilde{q}}(q_c)$ is the cumulative probability density of $\tilde{q}$ over $[q_c, \infty]$. In general, any cumulative probability density $F_X(Y)$ for continuous variables $X$ and $Y$ obeys a uniform distribution over $[0,1]$ if $Y$ obeys the same probability distribution as that of $X$ [1]. Therefore, under the null model, where $q_c$ and $\tilde{q}$ obey the same probability distribution, the $p$-value obeys the uniform distribution over $[0,1]$.

## II. DEPENDENCE OF THE STATISTICAL RESULTS ON THE NUMBER OF RANDOMISED NETWORKS

In this section, we examine the robustness of the statistical results with respect to the number of generated randomised networks. We use the 12 empirical networks used in the main text, which consist of different numbers of nodes and communities. For each community $c$, we compute the $p$-value using $R$ randomised networks, denoted by $p_c^{[R]}$. In the main text, we set $R = 500$. Then, we generate another $R$ randomised networks and compute the $p$-value, $p_c^{[2R]}$, for each community $c$. We measure the quality and the size of a community using $q_c^{\mathrm{mod}}$ and $\mathrm{vol}_c$, respectively. We use the Louvain algorithm to detect communities in the randomised networks.

The $p$-value computed with 500 randomised networks is close to that computed with $1,000$ randomised networks for most communities (Fig. S1(a)). The Pearson correlation coefficient, denoted by $r$, between the $p$-value between 500 networks and that with $1,000$

42networks is equal to $r = 0.999$. Additionally, the $p$-value with $R = 1,000$ is smaller than that with $R = 500$ for most communities, which indicates that the present statistical test is conservative when $R$ is small. Therefore, with $R = 500$ employed in the main text, which is relatively small, we are not overestimating the significance of the detected communities.

A large network and community may require many samples of randomised networks, $R$, for a reliable estimation of the $p$-value. To examine this possibility, we plot the variation in the $p$-value, defined by $|p_c^{[R]} - p_c^{[2R]}|$, for each community $c$ in Fig. S1(c). The variation of $p$-value tends to be small for large communities although the correlation between the variation and $\text{vol}_c$ is small ($r = -0.144$). A negative (albeit weak) correlation that we have found implies that a larger community requires a smaller value of $R$, which encourages the application of our statistical test to networks larger than those examined in the present article. Finally, we examine the robustness of the $p$-value with respect to the number of nodes in the network. To this end, for each empirical network, we average the variation, $|p_c^{[R]} - p_c^{[2R]}|$, over all communities in the network. The averaged variation is not strongly correlated with the number of nodes in the networks (Fig. S1(e); $r = 0.087$).

The results remain qualitatively the same when $R = 5,000$ (Figs. S1(b), S1(d) and S1(f)).

---

[1] P. Embrechts and M. Hofert, Math. Meth. Ope. Res. **77**, 423 (2013).

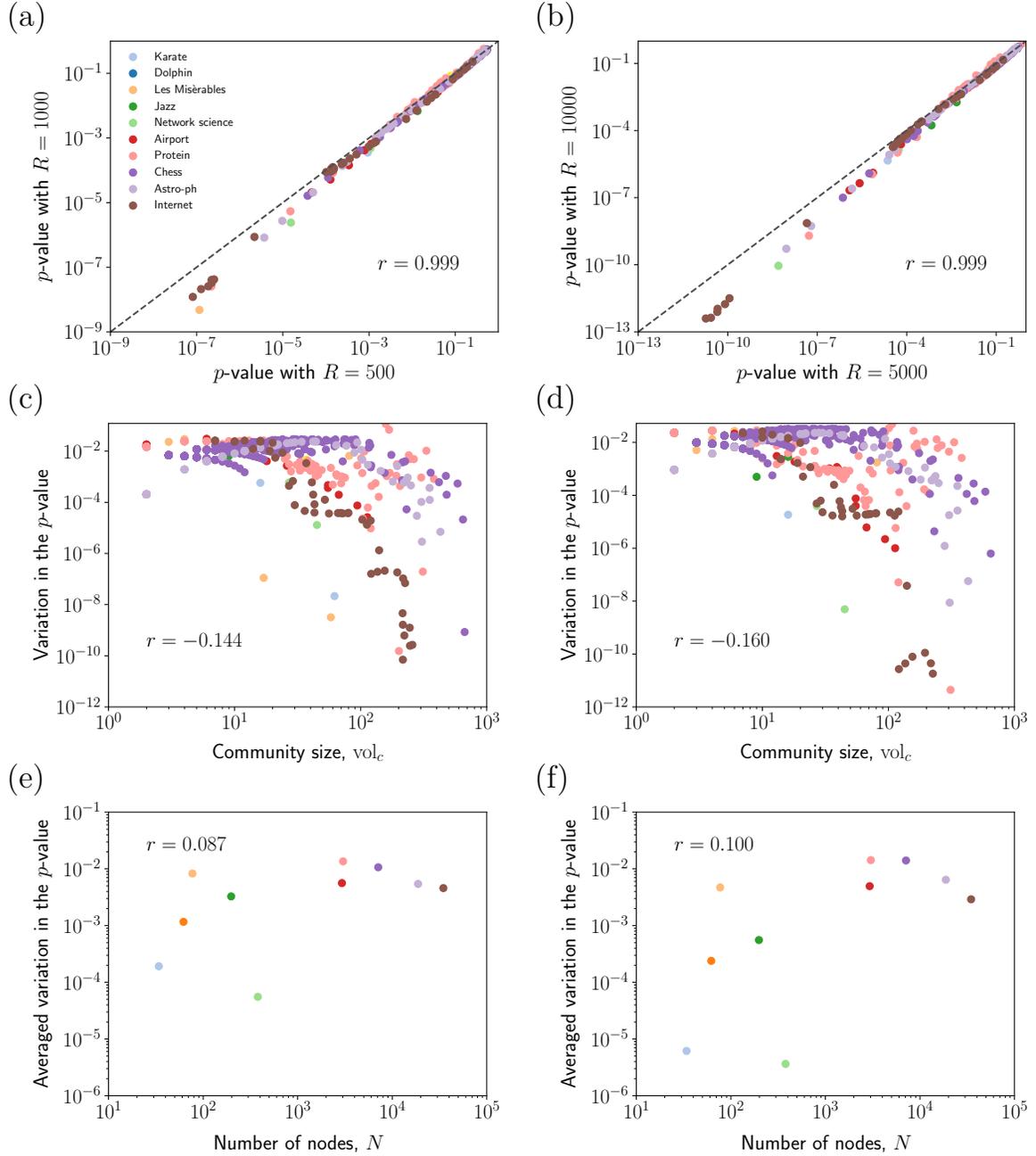

FIG. 1: Variation in the $p$-value computed with different numbers of randomised networks. Panels (a), (c) and (e) compare the results between $R = 500$ and $R = 1,000$. Panels (b), (d) and (f) compare the results between $R = 5,000$ and $R = 10,000$. In panels (a)–(d), each circle indicates a community detected in the original network. In panels (e) and (f), each circle indicates a network. We do not show the results for the Email and Blog networks because the $p$-value and their variation are less than $10^{-308}$. In each panel, $r$ denotes the Pearson correlation coefficient.



\end{document}